\documentclass[prl,twocolumn, aps,amssymb,footinbib,showpacs]{revtex4-1}

\usepackage{graphicx}
\usepackage{lipsum}
\usepackage{amsmath,amssymb}
\usepackage{color}

\newcommand{\ket}[1]{\ensuremath{\left| #1 \right\rangle}}
\newcommand{\bra}[1]{\ensuremath{\left\langle #1 \right|}}

\begin{document}

\title{Quasi Many-body Localization in Translation Invariant Systems}

\author{N. Y. Yao$^{1}$, C. R. Laumann$^{2}$, J. I. Cirac$^{3}$, M. D. Lukin$^{4}$, J. E. Moore$^{1}$}

\affiliation{$^{1}$Department of Physics, University of California, Berkeley, CA 94720, U.S.A.}
\affiliation{$^{2}$Department of Physics, University of Washington, Seattle, WA 98195, U.S.A.}
\affiliation{$^{3}$Max-Planck-Institut fur Quantenoptik, Hans-Kopfermann-Str.~1, D-85748 Garching, Germany}
\affiliation{$^{4}$Department of Physics, Harvard University, Cambridge, MA 02138, U.S.A.}

\begin{abstract}
 It is typically assumed that disorder is essential to realize Anderson localization. 
 Recently, a number of proposals have suggested that an interacting, translation invariant   system can also exhibit localization.
  We examine these claims in the context of a one-dimensional spin ladder. 
  At intermediate time scales, we find slow growth of entanglement entropy  consistent with the phenomenology of many-body localization.
  However, at longer times, all finite wavelength spin polarizations decay in a finite time, independent of system size.
  We identify a single length scale which parametrically controls both the eventual spin transport times and the divergence of the susceptibility to spin glass ordering.
  We dub this long pre-thermal dynamical behavior, intermediate between full localization and diffusion, quasi-many body localization.  
  %  identify a quasi-many body localized phase wherein
%
% We propose and analyze quasi-many body localization.
%
%  In such systems, an effective disorder landscape is generated internally via a separation of dynamical time-scales and localization manifests as an exponential susceptibility to translation-breaking perturbations. Here, we examine these claims for an interacting spin-ladder model. We observe eventual thermalization in all systems; however, a quasi-many-body localization plateau can exist as a finite-size effect. Entanglement entropies, decays of fractional polarization, and translation-breaking perturbations will be analyzed and correlated.
\end{abstract}
\maketitle

Since its proposal in 1958 \cite{Anderson58}, Anderson localization has been observed in disordered systems composed of photons, phonons, electrons and even ultracold atoms \cite{Wiersma97,Schwartz07,Kondov11}. 
The physics of localization in each of these systems can be largely understood as a single particle phenomenon. 
Extending disordered localization to the interacting many-body regime has attracted tremendous recent attention  \cite{Fleishman80,Altshuler97,Basko06,Gornyi05,Burin06,Oganesyan07,Pal10,Znidaric08,Monthus10,Bardarson12,Vosk12,Iyer13,Serbyn13,Huse13,Serbyn13b,Huse13b,Pekker13,Vosk13,Bahri13,Chandran13,Bauer13,Swingle13,Serbyn14,Vasseur14,Gopalakrishnan14,Nandkishore14a,Kjall14,Nandkishore14b,Agarwal14,Schiulaz13}, in part, because it represents a fundamental breakdown of quantum statistical mechanics. 
This breakdown opens the door to a number of possibilities, including novel phase transitions in high-energy states, the protection of quantum and topological orders, and even, the possibility of quantum information processing with disordered many-body systems \cite{Bahri13,Chandran13,Huse13b}.  

A number of recent proposals have investigated the possibility that localization can persist even in the absence of disorder \cite{Schiulaz13,Grover13,Roeck14,Hickey14}. 
This idea can be traced back to early work on $^3$He defects dissolved in solid $^4$He. 
There, Kagan and Maksimov \cite{Kagan74, Kagan84} proposed the intriguing possibility that a uniform system of strongly interacting narrow bandwidth particles could self-localize:~a subset of the $^3$He defects form immobile clusters which in turn block the diffusion of the remaining particles.

In more recent proposals, the distinction between mobile and immobile particles is imposed manually.
These models typically involve two types of particles, light and heavy; the dynamics of the heavy particles are significantly slower than those of the light particles \cite{Schiulaz13,Grover13}. At short time-scales, interactions between the two flavors serve as a random quasi-static background potential for the light particles. If strong enough, this effective disorder can localize the light particles and it has been argued that transport owing to the slow dynamics of the heavy particles is insufficient to delocalize the system. 
%Alternate models involving the motion of long-range, power-law interacting impurities or generalized Bose-Hubbard models have also been studied with similar conclusions. 
The central question which has emerged from these studies is whether randomness in the state of the system can be enough to cause ``self-localization'' and, what, precisely, does this mean?

\begin{figure}
	\centering
		\includegraphics[width=2.7in]{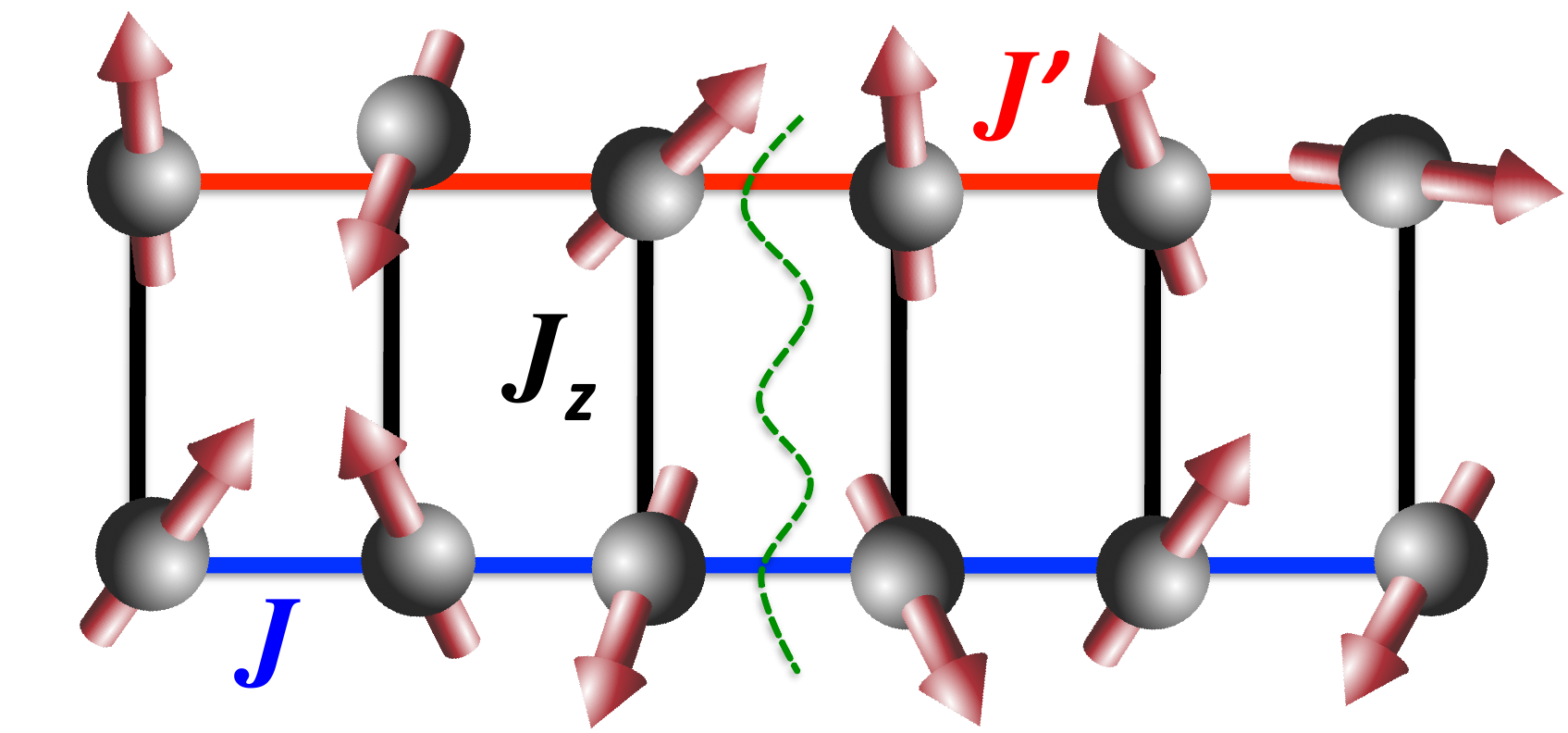}
	\caption{%
	Schematic of the one-dimensional spin-$1/2$ ladder.  $\sigma$-spins reside in the top chain and $S$-spins in the bottom. Along each chain, the spins are coupled by nearest neighbor XY (flip-flop) interactions with strength $J'$ and $J$ respectively. Across each rung, spins are coupled via Ising interactions of strength $J_z$. The green dotted line indicates the position of the cut used to divide the ladder when evaluating the entanglement entropy. }
	\label{fig:fig1_v1_schem}
\end{figure}

In this Letter, we address this question by considering a translation invariant spin ladder (Fig.~1), whose two legs carry, respectively, the fast and slow particles.
We find strong evidence for the existence of an exponentially diverging time-scale, $\tau \sim e^{L/\xi}$, which controls the decay of spin polarization at the longest available wavevector, $k = 2\pi/L$. 
We term this quasi-many-body localization (qMBL).
This should be contrasted with the behavior observed in \emph{disordered}, many-body localized spin chains where an initial polarization \emph{fails} to decay even at infinite time. 

For any finite wavevector $k$, we observe full polarization decay on a time-scale consistent with $\tau(k) \sim e^{1/(k \xi)}$, \emph{independent of system size}. Similar anomalous diffusion laws are seen in generalized Sinai models, where non-interacting particles diffuse in a random force-field \cite{Sinai82,Bouchaud90} and in the so-called spin-trapping of one-dimensional ferromagnetic Bose gases  \cite{Zvonarev07}. 
%
%This is reminiscent of anomalous diffusion for particles driven by a random correlated force. 
The same exponential divergence with length scale $\xi$  appears in the system's susceptibility to spin glass ordering \cite{Schiulaz13}. 
In previous work, this has been taken to imply an instability toward spontaneous many-body localization. However, the presence of anomalous diffusion rules out this scenario. 
In passing, we characterize several intermediate time-scale phenomena which are suggestive of ``self-localization'', but which ultimately give way to diffusion. 
%In previous work, this has been taken to imply an instability toward spontaneous many-body localization. 

Consider a two-leg, spin-$1/2$, ladder as shown in Fig.~1, with Hamiltonian,
\begin{align}
	\label{eq:hamgeneral0}
	H = \sum_{\langle ij \rangle} JS^+_i S^-_j+  \sum_{\langle ij \rangle} J' \sigma^+_i \sigma^-_j  + \sum_i J_z S^z_i \sigma^z_i + h.c
\end{align}
Spins of the lower (upper) chain are labeled $S$ ($\sigma$) and are coupled via a nearest neighbor $XY$ interaction of strength $J$ ($J'$). The two spin species are coupled across a rung via Ising interactions of strength $J_z$. In the limit, $J' \rightarrow 0$, the $\sigma$ spins of the upper chain can be viewed as classical variables that generate quenched disorder for their $S$-spin cousins. In this limit, fermionization of the $S$-chain produces a non-interacting model which localizes for typical configurations of the $\{ \sigma_i \}$.
%
%Depending on the state of the $\{ \sigma_i \}$, the lower chain sees an effective random on-site field of strength $ \pm J_z$, which induces single particle localization of the $S$ chain. 
%
The introduction of a finite $J'$ drives dynamics in the $\sigma$ chain and effectively induces interactions in the system. Thus, perturbative $J'$ are the limit in which one might hope to observe the strongest signatures of localization. 
%
%
%Formally, this arises because strings of the Jordan-Wigner transformation no longer cancel between all pairs of sites. 
%
%Thus, even a perturbative $J'$ effectively induces interactions in the system, and the question becomes: does the system transition from single-particle to many-body localized and if so, does localization persist to infinite times. 

\begin{figure}
\hspace{-5mm}\includegraphics[width=3.6in]{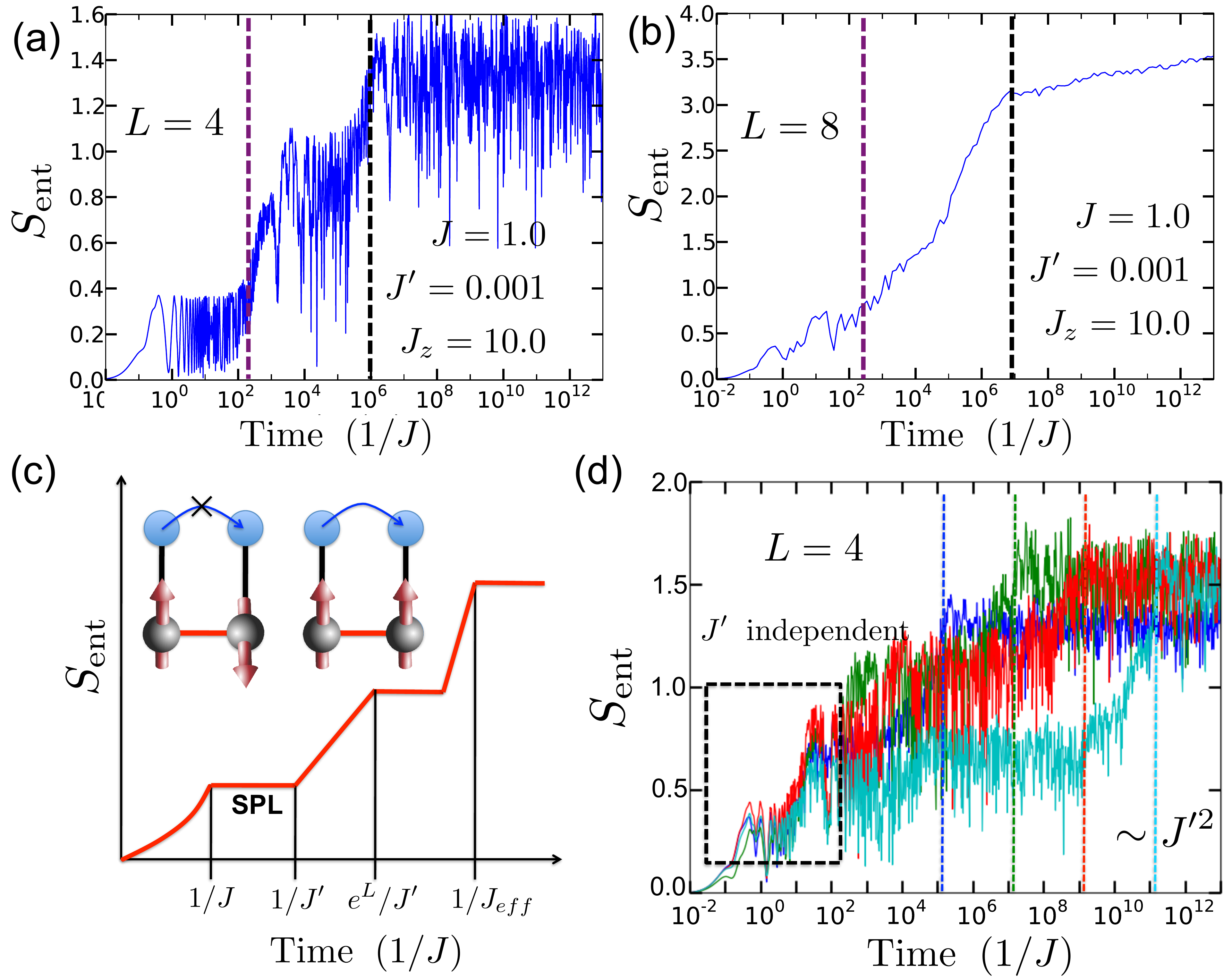}
\caption{(a) Growth of entanglement entropy for $L=4$ sites, with $J' = 10^{-3}$ and $J_z=10$. The entanglement entropy is averaged over 30 random initial product states. (b) Analogous data for $L=8$ averaged over 100 random initial product states. (c) Schematic  short-time entanglement entropy behavior. At $t_1 \sim 1/J$ a single particle localized (SPL) plateau is observed. At time-scales $\sim  1/J'$ interactions set in and a logarithmic growth of entanglement begins. At $t_2 \sim e^L/J'$ this  growth  saturates for short chains unless it is pre-empted by the final ``plateau''  at $t_3 \sim 1/J_{eff} = (J'^2/J_z)^{-1}$ \cite{suppinfo}. (d)  $S_{ent} (t)$ for Heisenberg coupling along the ladders with $L=4$, $J_z=10$, $J' = 10^{-2}, 10^{-3}, 10^{-4}, 10^{-5}$.  Since there is no single-particle limit, the initial dynamics are $J'$ independent, but $t_3$ continues to scale as $J'^2$.
 }
\label{fig:mscheme}
\end{figure}

\emph{Entanglement dynamics}---
%In this section, we consider the growth of entanglement entropy over 
%
We perform extensive exact diagonalization studies of  Eq.~\eqref{eq:hamgeneral0}. We consider periodic systems up to $N=2L = 24$ sites and work at fixed filling $\nu_{s/\sigma} = 1/2$, where $\nu$ is the fraction of $S/\sigma$-sites with spin-up divided by $L$.
 All energies are normalized to $J=1$. 
Our first diagnostic is the growth of entanglement entropy $S_{ent} = -\text{tr} \rho_A \log \rho_A$ across a central cut (parallel to a rung, Fig.~1) that divides the system in sub-regions $A$ and $B$. Initial states are chosen to be random product states within the relevant Hilbert space and we average over  $30$-$100$ states depending on system size. 
For short and intermediate time-scales, much of the observed entanglement dynamics can be understood within the framework   of spontaneous MBL \cite{Schiulaz13}. As we will see, it is only at the longest time-scales that this framework fails and anomalous diffusion sets in.

We begin with strong effective disorder, $J_z=10$, and measure the effect of a small $J' \ll J$.
The result for $L=4$, $J' = 10^{-3}$ is shown in Fig.~2a (see \cite{suppinfo} for $J' = 10^{-2}, 10^{-4}$).   We observe three plateaus in the growth of the entanglement entropy, which can be qualitatively understood as follows (Fig.~2c).
There is an initial growth of $S_{ent}$ until time $t_1 \sim 1/J$, arising from the rapid expansion of wave packets to a size of order the non-interacting localization length. The first plateau is consistent with the entanglement behavior for single particle localized states and persists indefinitely for $J'=0$.

At time-scale $t \sim 1/J'$ (purple dashed lines, Fig.~2), logarithmic growth of entanglement sets in \cite{Bardarson12,Serbyn13},  indicating that $J'$  sets the time-scale of interactions. The relevant dephasing process is shown in the inset of Fig.~2c: the  single-particle states of the $S$-chain experience an energy shift dependent on their local occupation.  This effective density-density interaction results from the hybridization of  $\sigma$-chain orbitals on the time-scale $J'$.
Logarithmic growth progresses until the entanglement saturates at a second plateau, $t_2 \sim e^L/J'$. Up to now, the entanglement dynamics are consistent with those observed in disordered, MBL systems. 

The second plateau corresponds to the complete, finite-size, dephasing of the $S$-chain, while the dephasing dynamics of the slower $\sigma$-chain have yet to begin. In principle, one might expect the third plateau to correspond to the finite-size entanglement saturation of the full system, with $t_3$ also scaling as $\sim e^L$.
However, by numerically varying $J'$, $J_z$, and $L$ we find instead that  $t_3 \sim (J'^2/J_z)^{-1}$, with a weak, sub-exponential, $L$-dependence (black dashed lines, Fig.~2a,b) \cite{suppinfo}. 
At larger system sizes (Fig.~2b), the intermediate plateau  disappears since $t_2 \sim e^L$ while $t_3$ does not;
moreover, the third ``plateau'' begins to exhibit a clear upward drift, presaging additional dynamics to come. 

This picture of entanglement growth is further confirmed by generalizing Eq.~\eqref{eq:hamgeneral0} to Heisenberg couplings within each chain. In particular, there is no longer  a non-interacting regime as is evidenced by the $J'$ independent short-time dynamics in Fig.~2d; meanwhile, the final ``plateau'' continues to depend quadratically on $J'$ with weak $L$-dependence. We return to the $XY$ model for the remainder of the manuscript. 

% TODO Where does this go if anywhere? J_{eff} why physically is this the right time scale -- two possibilities. 

%
% To understand the microscopic origin of these various time-scales we consider the scaling of $t_{int}$ and $t_d$ as a function of $J'$. Comparing the growth of $S_{ent}$ across the various parameters yields $t_{int} \sim 1/J'$ and $t_{int} \sim 1/J'^2$; similarly, holding $J, J'$ fixed while varying $J_z$ yields that $t_{int} \sim J_z$ (see supplementary information). The scaling of $t_{int}$ is consistent with a picture of $J'$ induced departure from single-particle localization. In particular, the effective induced-interaction-strength  is directly proportional to $J'$ and arises from the delocalization of upper chain particles across a bond where a pair of $S$ spins are aligned (inset Fig.~3a). The scaling of $t_d \sim J_z / J'^2$ suggests that the eventual diffusion in the upper chain is characterized by a diffusion constant $\sim J'^2/J_z$ consistent with a picture of off-resonant hopping across a flipped $S$-rung (Fig.~3b). \textcolor{red}{Process $J J' / J_z$?} The decay of fractional polarization confirms that the final plateau correlates with diffusion of the slow, upper chain as $t_d$ corresponds to the location where the upper chain $D_{k=1}$ reaches its final decay step. Interestingly, it also corresponds to the location where the $k=2$ polarization first decays to zero.

\emph{Long-time dynamics}---
To understand the long-time dynamics, it is helpful to turn to other physical quantities. 
In particular, we probe the  decay of  spin  polarization as well as the susceptibility, $\chi$ to spin glass ordering. 
As we will see, a \emph{single} length scale, $\xi$, controls both as we vary $J_z$ and $\nu_s$ (holding $\nu_\sigma = 1/2$, $J' =0.01$). Indeed,  the time-scale $\tau$ for ultimate polarization decay scales as $\sim e^{1/(k\xi)}$, while the susceptibility scales as $\sim e^{c L/\xi}$, for a constant $c$. 
In the $J'\to0$ limit, $J_z$ and $\nu_s$ directly control the effective disorder and thus the localization length $\xi_0$.  From the observed behavior of $\xi (J_z, \nu_s)$, we surmise that it is continuously connected to $\xi_0$ as one turns off $J'$.

\begin{figure}
\includegraphics[width=3.4in]{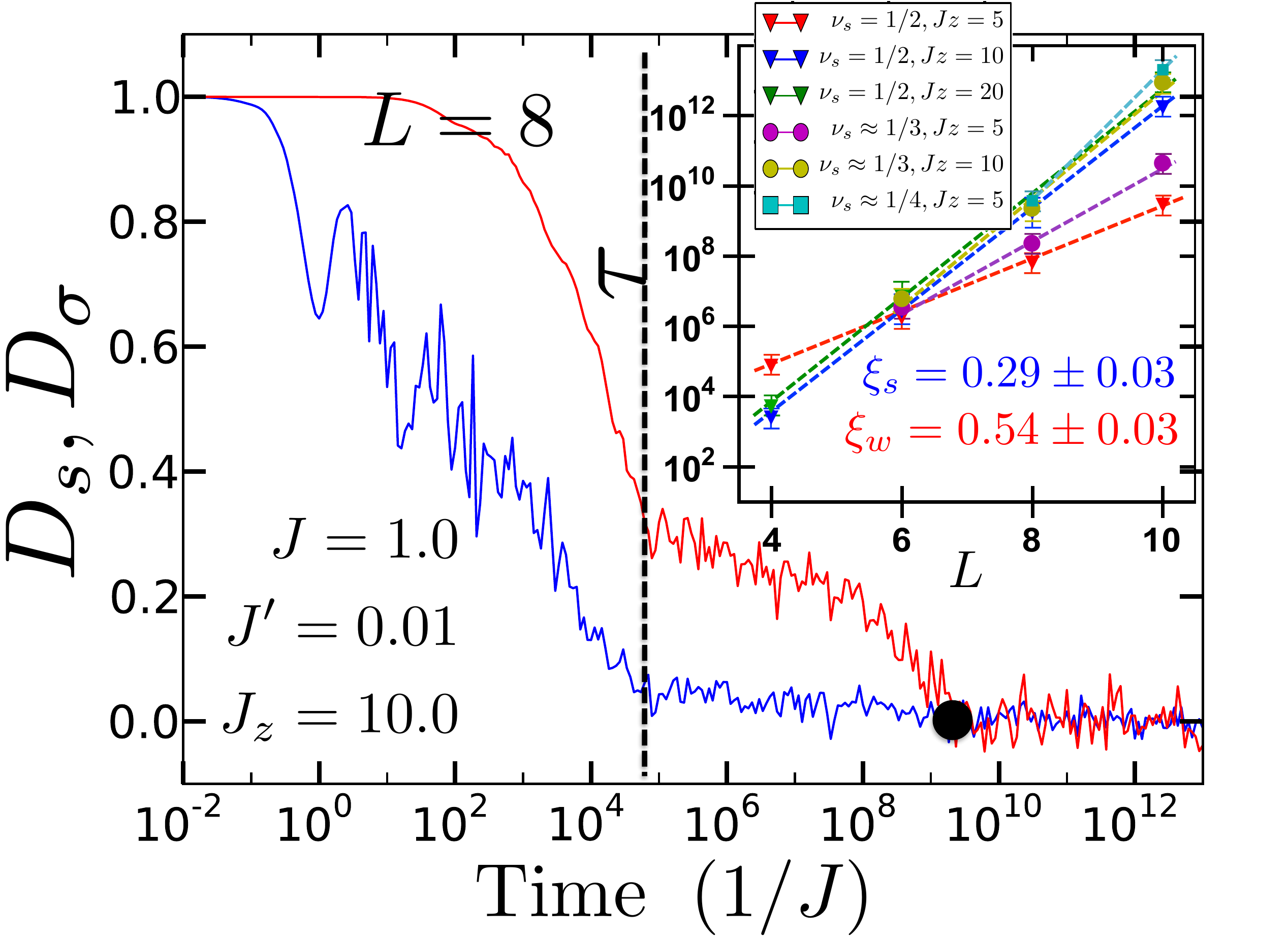}
\caption{  A typical time trace of the decay of spin polarization. The blue line shows $D_s(t)$ while the red line shows $D_\sigma (t)$. The black dashed line indicates the position of $t_3$ as determined from the entanglement entropy and the black dot depicts the time at which all polarization has fully decayed. 
(inset) Depicts $\tau (k=2\pi/L)$ as a function of system size  for $J_z=5,10,20$ and filling fractions $\nu_s=1/2$, $\nu_s \approx 1/3,1/4$.
For $\nu_s \approx 1/3, 1/4$ commensuration effects at finite size prevent the choice of identical filling across sizes. For $\nu_s\approx1/3$, the sizes are $L=6,8,10$ with  $\nu_s=1/2,3/8,3/10$. For $\nu_s \approx 1/4$, the sizes are $N = 8,10$ with $\nu_s = 1/4,1/5$ respectively \cite{suppinfo}.
The data at $\nu_s \approx 1/3,1/4$ are qualitatively consistent with saturation at  strong effective disorder. 
%
%prevent a clear quantitative analysis.
%
%
%%We note that the $\nu=1/3$ data are qualitatively consistent but the commensuration effects at finite size prevent a clear quantitative analysis. 
% the time-scale for spin polarization decay (at wavevector $k=2\pi/L$), as a function system size $L$. Data are obtained at different $J_z=5,10,20$ and filling fraction $\nu=1/2,1/3,1/4$ (of the $S$-chain), where $\nu$ is defined as the fraction of sites with spin-up divided by the length. Note that the filling fraction of the $\sigma$-chain is always $1/2$.
%%
% For $\nu=1/2$, the system sizes are $N=8,12,16,20$. For $\nu=1/3$, the sizes are $N=12,16,20$ with  $n_s=2,3,3$ spin-ups in the $S$-chain respectively. For $\nu=1/4$, the sizes are $N = 16,20,24$ with $n_s = 2,2,3$ respectively. Data at the largest sizes $L=20,24$ are obtained via shift-and-invert Lanczos (SIL) \cite{suppinfo}. 
%
  }
\label{fig:mscheme}
\end{figure}

\begin{figure}
\includegraphics[width=2.8in]{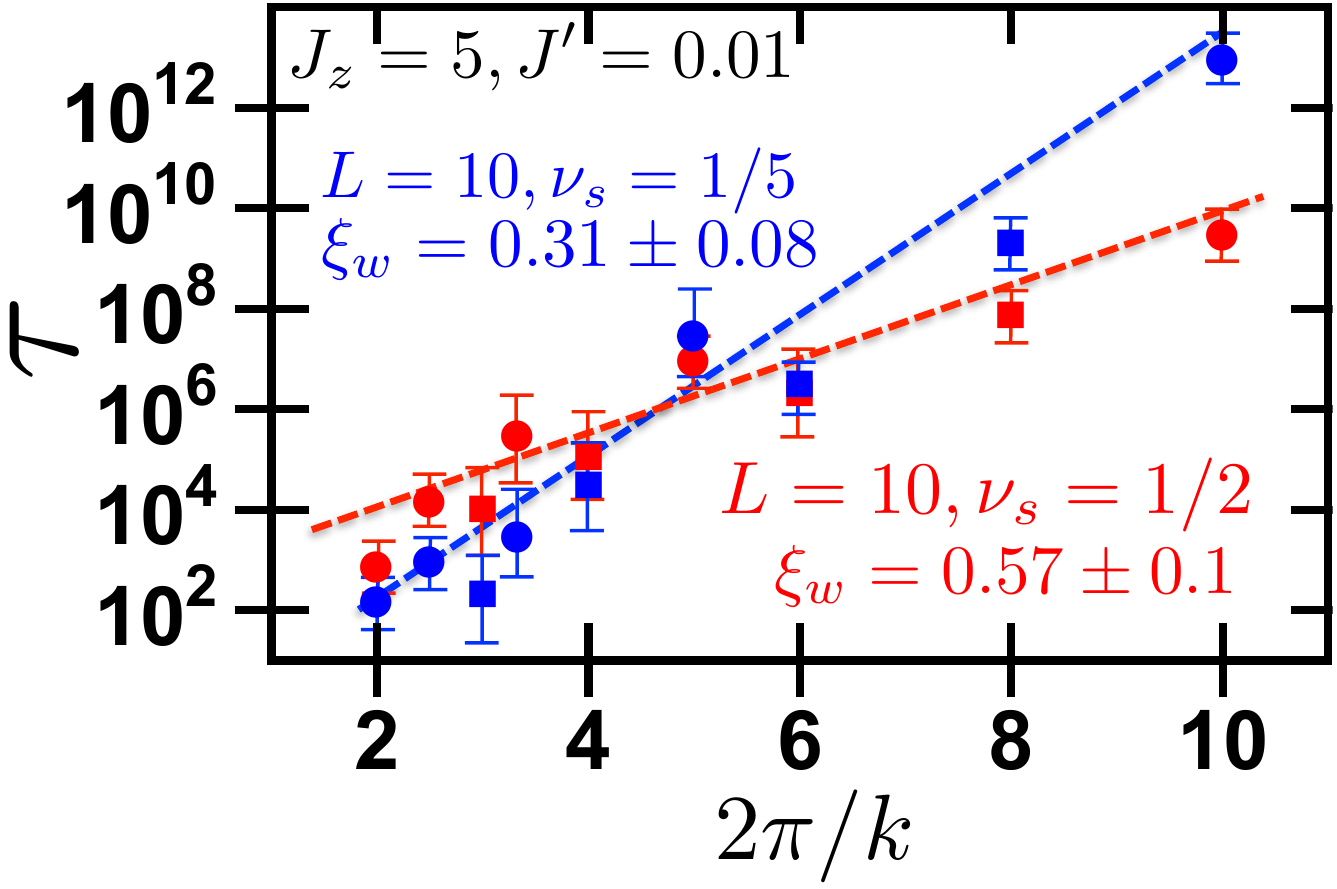}
\caption{ Depicts $\tau$ as a function $k$. Data are obtained at $J_z=5$, $J'= 0.01$. Red circles correspond to $L=10, \nu_s=1/2$ while blue circles correspond to $L=10, \nu_s=1/5$. 
To rule out additional dependencies, a few different $k$ are also shown for $L=6,8$ (square data). 
  }
\label{fig:mscheme}
\end{figure}

%define quasil MBL and thermalization better. 

 %Interestingly, this same length scale also controls the susceptibility to spin glass ordering. 

\emph{Polarization decay}---The decay of polarization $D_{s}$ ($D_\sigma$) is a measure of spin transport at infinite temperature \cite{Pal10}.
As each flavor of spin is separately conserved, we perturb the system with a small inhomogeneous spin modulation of the form $\hat{F_s} (k) = \sum_j S^z_j e^{i  k j}$ (similarly for $\hat{F}_\sigma$) and measure the time-dependent relaxation of this polarization,
\begin{align}
  \label{eq:decayseries}
  D_{s/\sigma}(k,t) = \left\langle e^{-i H t} F^\dagger_{s/\sigma} (k)  e^{i H t} F_{s/\sigma} (k) \right\rangle
\end{align}
where the average is taken at infinite temperature.
% For each eigenstate $k$, the dynamic polarization is given by
% \begin{equation}
% D_{s/\sigma}^k = 1 - \frac{ \langle k | \hat{F}_{s/\sigma}^{\dagger} | k \rangle \langle k| \hat{F}_{s/\sigma} | k \rangle}{ \langle k | \hat{F}_{s/\sigma}^{\dagger} \hat{F}_{s/\sigma} | k \rangle},
% \end{equation}
% and we define $D_{s/\sigma}$ as the infinite temperature average of $D^k_{s/\sigma}$.

A typical time-series for $D_{s/\sigma}$ at the longest wavelength $k=2\pi/L$ is depicted in Figure 3. The most remarkable feature is the clarity of the various time-scales. For example, the dashed line indicates the timescale $t_3$ as extracted from the corresponding entanglement entropy \cite{suppinfo}. 
At this point, it is clear that there is still significant residual polarization.  However, it is also clear, that this polarization fully decays by time $\tau \sim 10^9 /J$. 
The parametric dependence of $\tau (k=2\pi/L)$ is illustrated in the inset of Fig.~3, where we plot its $L$-dependence at fixed $J_z$ and $\nu_s$; $\tau$ scales exponentially in system size which defines the length scale $\xi (J_z, \nu_s)$ as the inverse slope of the curves. 
For weak effective disorder $\nu_s =1/2, J_z = 5$, we find $\xi_w = 0.54 \pm 0.05$. All other parameters correspond to stronger disorder, producing a shorter length, which saturates at $\xi_s = 0.29 \pm 0.03$ (extracted from $\nu_s =1/2, J_z = 20$). 

The existence of the length-scale  $\xi$  suggests that finite wavelength inhomogeneities decay on a finite time-scale $\tau (k) \sim e^{1/(k\xi)}$. To test this hypothesis, we consider the ultimate decay time as a function of $k$ for fixed $L$. For $L=10$, $J_z=5$, $\nu_s = 1/2, 1/5$ this data (circles) is plotted in Figure 4 and is consistent with the proposed functional form (dashed line). 
This provides an independent means to extract $\xi$. We obtain $\xi_w = 0.57 \pm 0.1$ and $\xi_s = 0.31 \pm 0.08$ in agreement with the two lengths quoted above. 
To ensure that there is no system-size dependence lurking, we also plot  $\tau(k)$ for $L=6,8$ (squares). 

The behavior $\tau (k) \sim e^{1/(k\xi)}$ contrasts with both many-body localization, where inhomogeneities  never decay, and with  diffusion, where they decay as $\sim 1/k^2$.
Thus, the term  quasi-MBL.

\emph{Susceptibility}---Finally, following \cite{Schiulaz13}, we probe our system's susceptibility to spin glass ordering by introducing a perturbation of the form
\begin{align}
	\label{eq:hamgeneral}
	H_W = \sum_i h^z_i S^z_i  +  \sum_i h'^z_i  \sigma^z_i 
\end{align}
where $h,h'$ are independent random fields drawn from a uniform distribution of width $W$. To quantify the system's response to $H_W$, we consider an observable $\Delta \rho_\psi = \frac{1}{N} \sum_i^N | \langle \psi | S^z_{i+1} -  S^z_{i} |\psi \rangle | $ which measures the inhomogeneity of the spin polarization in the $S$-chain within an eigenstate $\psi$.
We perform exact diagonalization on $H_T = H + H_W$ with $\nu_s = \nu_\sigma = 1/2$, $J' = 0.01$, $J_z = 5,10,20,40$ and  $ 10^{-6}<W < 10^{-4}$. We average over $10^3$ disorder realization for $N=8,12$ and over $10^2$ realizations for $N=16$; we also average $\Delta \rho$ over 10 eigenstates centered around energy density $J/4$. Our results are depicted in Figure 5. 
The inset indicates that $\rho(W)$ is in the linear response regime as all data lie at slope one in the log-log plot.

It has been argued \cite{Schiulaz13} that an exponential in system size divergence of  $\chi = d \rho/ dW$ reflects an instability toward many-body localization. We indeed observe such a dependence (Fig.~5).  
However, as previously discussed, we do not view the system as truly MBL, since spin transport occurs, albeit slowly, across the full system. 
In fact, the transport time appears to be precisely correlated with the divergence of the spin glass susceptibility. 
An analysis of the exponential dependence of $\chi$ also yields an effective length scale as a function of $J_z$ (Fig.~5). 
This length is in fact proportional to $\xi$, with a proportionality factor $c \approx 2.6 \pm 0.1$ across the data \cite{suppinfo}. 
%
%We note that the $\nu=1/3$ data are qualitatively consistent but the commensuration effects at finite size prevent a clear quantitative analysis. 

\emph{Discussion}---
For a finite size system, translation invariance requires that at infinite time, any finite wavelength polarization must decay to zero. This follows immediately from Eq.~\eqref{eq:decayseries} after inserting a resolution of the identity and dephasing off-diagonal matrix elements \cite{footnote}: 
\begin{align}
  D_{s/\sigma}(\infty) &= \sum_\psi \bra{\psi} F^\dagger_{s/\sigma} \ket{\psi}\bra{\psi} F_{s/\sigma} \ket{\psi}.
\end{align}
As $F_{s/\sigma}$ carries non-zero momentum its diagonal matrix elements vanish between translation-invariant many-body eigenstates, $\ket{\psi}$.
Thus, although disordered MBL systems exhibit finite residual polarization, we cannot expect that of any translation-invariant system.
This algebraic truth does not rule out the possibility that the decay time, $\tau(k)$, of finite wavelength polarization diverges with the system size.
This is the natural definition of translation-invariant many-body localization.

While numerically accessible system sizes prohibit a complete characterization, we do not believe that such behavior holds. 
Rather, we find a finite decay time $\tau(k)$ for all $k$. 
All decay times, as well as the spin glass susceptibility $\chi$, are controlled by a single physical length scale $\xi$.
Crucially, this length scale is not simply related to the many-body density of states (inverse entropy).
As previously discussed, we surmise that $\xi$ is connected to the true localization length in the $J'\to0$ limit, despite the fact that the system is not localized for any $J' \neq 0$.

\begin{figure}
	\centering
		\includegraphics[width=3.0in]{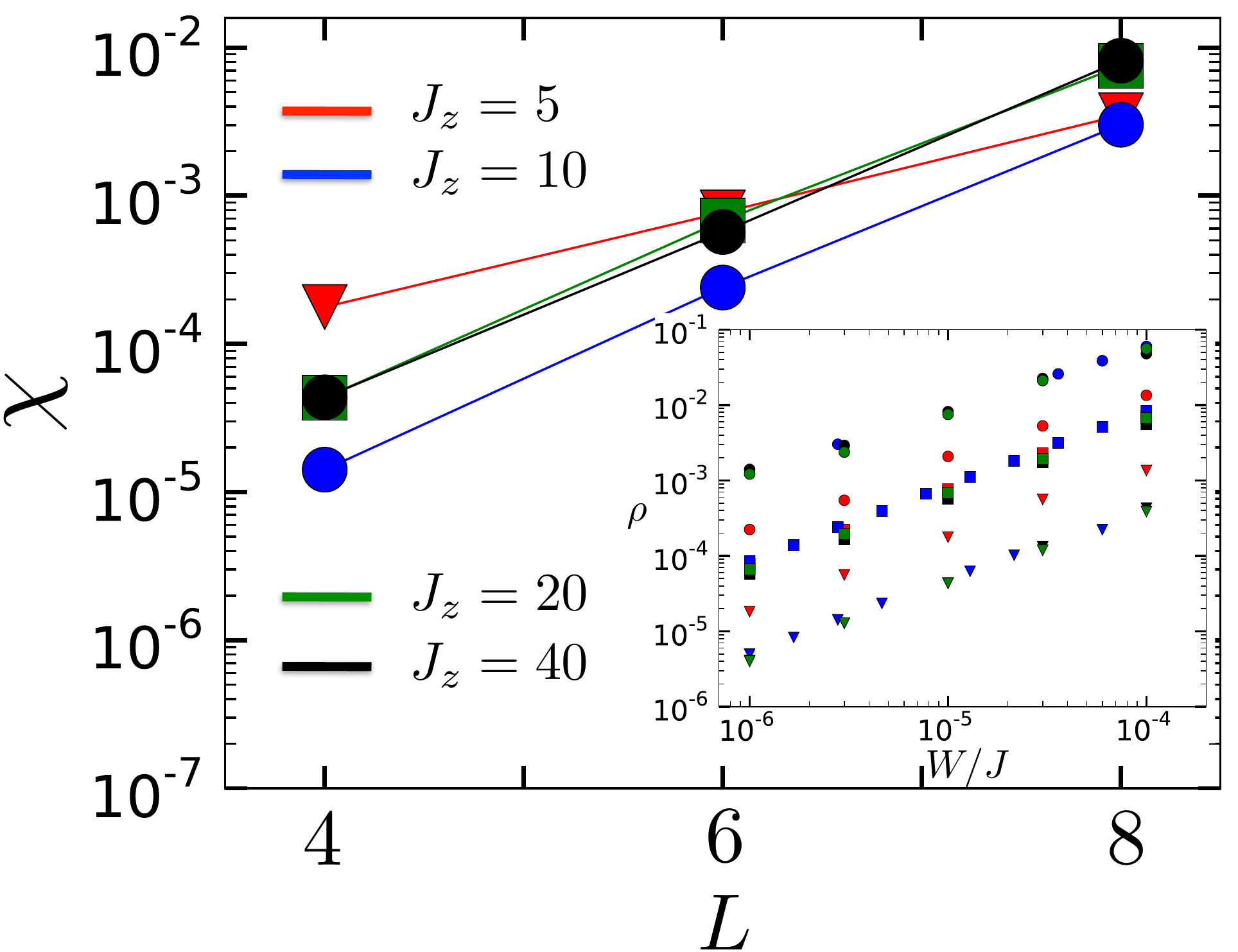}
	\caption{%
	(a) Depicts $\chi$, the susceptibility to spin glass ordering as a function of system size $L$ for $\nu=1/2$ and $J_z = 5,10,20,40$. (inset) Shows the raw $\rho (W)$ data used to generate the main figure. The value of $\chi$ is taken to be that of $\rho$ at $W=10^{-6}$. The slope of unity on the log-log plot demonstrates that we are clearly in the  linear response regime. 
  }
	\label{fig:fig1_v1_schem}
\end{figure}

In summary, by examining entanglement growth, the decay of  polarization, and the susceptibility to spin glass ordering, we provide evidence that translation invariant systems can exhibit quasi-MBL behavior intermediate  between full localization and diffusion. 
The entanglement dynamics are consistent with MBL-type growth at short and intermediate times, but ultimately give way to anomalous diffusion.
This behavior is characterized by polarization decay on a time-scale $\tau(k) \sim e^{1/(k \xi)}$, which in real space, corresponds to an anomalous random walk  with a mean square deviation growing as the log-squared of time. 
Such diffusion is reminiscent of spin-trapping in a one-dimensional ferromagnetic Bose gas  \cite{Zvonarev07}, as well as generalized Sinai diffusion models \cite{Sinai82,Bouchaud90}. It may also have qualitative similarities to the behavior observed in pre-thermalizing 1D multi-component bosons \cite{Kitagawa11,Gring12}. 
Both the long-time dynamics and the system's ``spin-glass'' susceptibility  are governed by the same physical length scale $\xi$, which reduces to the localization length in the $J' \to 0$ limit.

It is a pleasure to gratefully acknowledge the insights of and discussions with Z. Papic, S. Gopalakrishnan, M. Knap, D. Huse, A. Chandran, S. Choi, S. L. Sondhi. This work was supported, in part, by  DMR-1206515, the  Miller Institute for Basic Research in Science,  the NSF, the AFOSR-MURI.  CRL acknowledges the hospitality of the Perimeter Institute and the Simons Institute for the Theory of Computing. 

During the completion of this work, a related preprint has appeared \cite{Schiulaz14}.


\begin{thebibliography}{99}
\bibitem{footnote}  We assume that there are no additional lattice symmetries, such as reflection, which protect degeneracies.
\end{thebibliography}
\end{document}

\title{Supplemental Material for Translation Invariant Quasi Many-body Localization}
\author{N. Y. Yao, C. R. Laumann, J. I. Cirac, M. D. Lukin, J. E. Moore}
\maketitle

\begin{figure*}[ht]
\begin{center}
\includegraphics[width=0.8\textwidth]{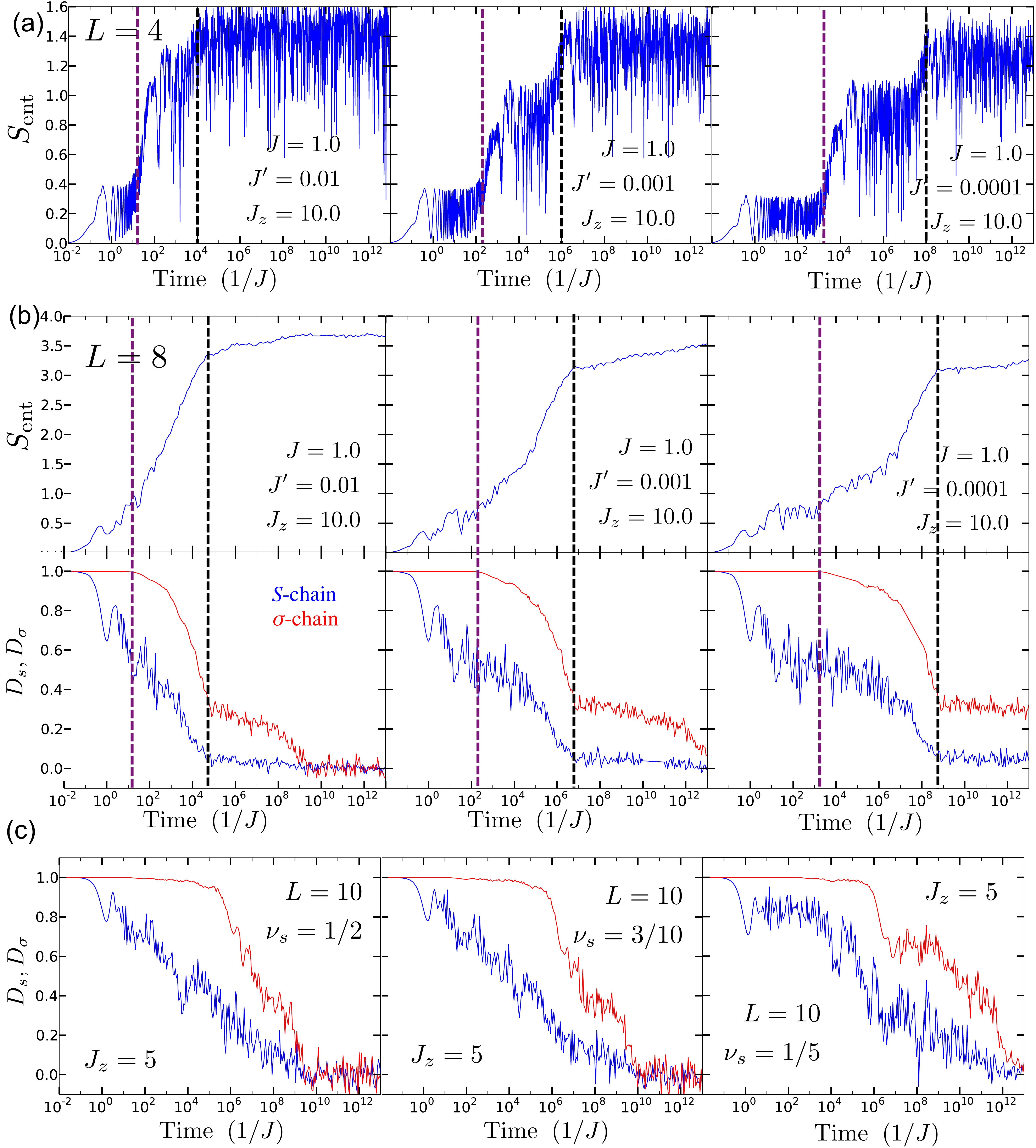}
\end{center}
\caption{ Entanglement numerics are performed at $\nu_s = \nu_\sigma = 1/2$. (a) Growth of entanglement entropy for $L=4$ sites, with $J' = 10^{-2},10^{-3},10^{-4}$ and $J_z=10$. The entanglement entropy is averaged over 30 random initial product states. 
(b) 
Upper panel: Analogous data for $L=8$ averaged over 100 random initial product states. 
Lower panel: Shows polarization decay at $k=2\pi/L$ for the $S$ and $\sigma$-chains at infinite temperature. The purple dashed lines correspond to the beginning of the log growth of entanglement after the first `single particle' plateau. The black dashed lines correspond to $t_3$, the beginning of the final phase of polarization decay.
 (c) Depicts polarization decay data for $L=10$ at different filling fractions $\nu_s= 1/2, 3/10,1/5$ of the $S$-chain.
 }
\label{fig:mscheme}
\end{figure*}

\section{Numerical data and details}

Figure S1(a) shows the data for entanglement entropy growth for $L=4$ with $J' = 10^{-2},  10^{-3},  10^{-4}$.
The time-scale for the first logarithmic growth of $S_{ent}$ scales linearly with $1/J'$ while $t_3 \sim J'^2$. Figure S2(b) depicts the analogous data for $L=8$ in the upper panel and correlates it with the decay of long wavelength polarization in the bottom panel. 
The final stage of the dynamics, beginning at $t_3$ exhibits very slow entanglement growth while there is still significant residual polarization.  However, at even later times, all residual polarization decays to zero. 
Figure S1(c) shows $D_s(t)$, $D_\sigma(t)$ as a function of decreasing filling fraction $\nu_s$ in the $S$-chain. The time-scale for full polarization decay is significantly increased for smaller $\nu_s$, consistent with the simple picture that the amount of effective disorder seen by the $S$-spins increases at smaller filling fraction. 

Data at $L=10$ and $12$ are obtained via shift-and-invert Lanczos. To ensure that the data has converged, we progressively increase the number of eigenstates averaged over from $100$ to $500$ and check convergence of the time traces. We also perform shift-and-invert centered at different energy densities, $J/2$ and $J/4$ and find that both give quantitatively similar time traces. 

  \begin{figure*}
\begin{center}
\includegraphics[width=0.6\textwidth]{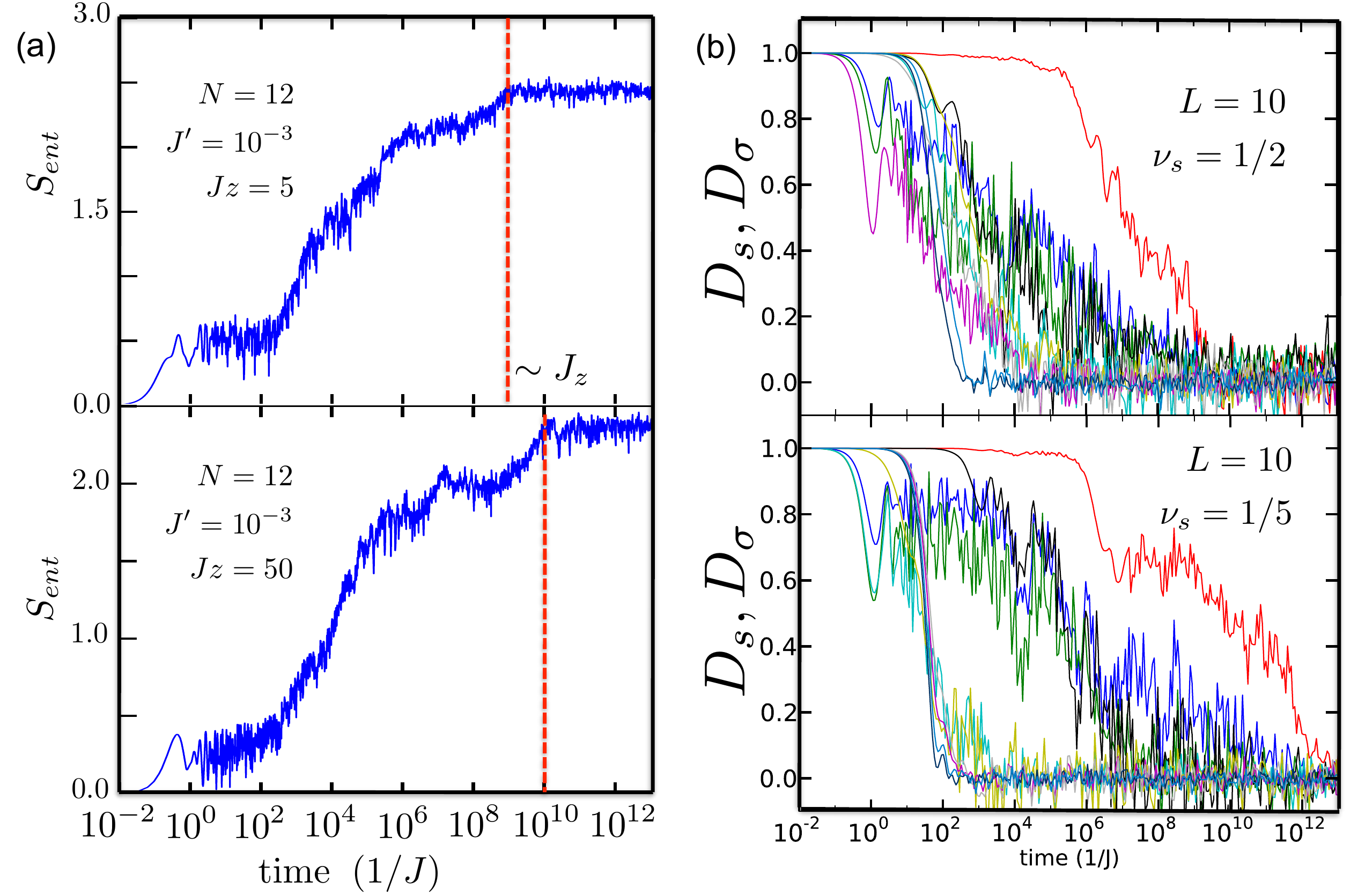}
\end{center}
\caption{ (a) Representative data showing growth of entanglment entropy for $L=6$ with fixed $J'$ and varying $J_z$. From the quadratic scaling of $t_3$ with $J'$ shown in Fig.~S1 and the linear scaling with $J_z$ shown here, one finds $t_3 \sim 1/J_{eff}$ with $J_{eff} = J'^2/J_z$. (b) Decay of polarization for different $k$ at $L=10$, $\nu_s = 1/2, 1/5$.  }
\label{fig:mscheme}
\end{figure*}

 Figure S2(a) shows representative data at $L=6$ from which we can read off the scaling of the third plateau with $J_z$ and confirms that $J_{eff} = J'^2/J_z$.  Figure S2(b) shows the raw data for different $k$ at $L=10$, $\nu_s = 1/2, 1/5$.
In order to numerically extract parametric dependences of the observed time scales, we have collected polarization, entanglement and susceptibility data at sizes $L=4,6,8,10,12$, fillings of the S-chain $\nu_s=1/2$, $\nu_s \approx 1/3,1/4$, fillings of the $\sigma$-chain $\nu_\sigma=1/2$, and couplings varying $J' = 10^{-5}, 10^{-4}, 10^{-3}, 10^{-2}$ and $J_z = 5,10,20,40,50,100$, of which only a few representative traces are included in the main text or supplementary information in order to contribute to the fight against visual spam.